\documentclass[twocolumn]{aastex63}
\pdfoutput=1

\usepackage{mathrsfs}

\def\lens{{\textsc{Lenstronomy}}}

\usepackage{amsmath}
\usepackage{enumitem}
\usepackage{hyperref}
\hypersetup{colorlinks,linkcolor={blue},citecolor={blue},urlcolor={blue}} 

\def\mbh{\textit{$\mathcal{M}_{\rm BH}$}}
\def\m{\textit{$\mathcal{M}_{\star}$}}

\def\mbulge{\textit{$\mathcal{M}_{\rm bulge}$}}
\def\logm{\textit{${\rm log}\,\m$}}
\def\logmbh{\textit{${\rm log}\,\mbh$}}
\def\logmbulge{\textit{${\rm log}\,\mbulge$}}

\def\deltalogmbh{\textit{$\Delta\,\logmbh$}}

\def\rel{\textit{$\mbh-\m$}}
\def\bulgerel{\textit{$\mbh-\mbulge$}}

\def\hb{\textit{${\rm H}\beta$}}

\def\lens{{\textsc{Lenstronomy}}}

\def\g{$g$}
\def\r{$r$}
\def\i{$i$}
\def\z{$z$}
\def\y{$y$}

\def\sid{\textit{$\rm S\acute{e}rsic\ index$}}

\def\ss{\textit{$\rm S\acute{e}rsic$}}

\def\mgii{{\rm Mg\ \scriptsize{II}}}

\def\msun{\textit{$M_\odot$}}
\def\ergs{\textit{$\rm erg\ s^{-1}$}}
\def\kms{\textit{$\rm km\ s^{-1}$}}

\def\gm{\textit{$\gamma$}}

\def\mvir{\textit{$m_{\rm vir}$}}

\def\lbol{\textit{$L_{\rm bol}$}}
\def\sigmamu{\textit{$\sigma_\mu$}}
\def\nuw{\textit{$\nu_{\rm FWHM}$}}
\def\edd{\textit{$\lambda_{\rm Edd}$}}
\def\loglbol{\textit{${\rm log}\,L_{\rm bol}$}}

\graphicspath{{./}{figures/}}

\begin{document}

\title{Synchronized Co-evolution between Supermassive Black Holes and Galaxies Over the Last Seven Billion Years as Revealed by the Hyper Suprime-Cam Subaru Strategic Program}

\author{Junyao Li}
\affiliation{CAS Key Laboratory for Research in Galaxies and Cosmology, Department of Astronomy, University of Science and Technology of China, Hefei 230026, China}
\affiliation{Kavli Institute for the Physics and Mathematics of the Universe, The University of Tokyo, Kashiwa, Japan 277-8583 (Kavli IPMU, WPI)}
\affiliation{School of Astronomy and Space Science, University of Science and Technology of China, Hefei 230026, China}

\author{John D. Silverman}
\affiliation{Kavli Institute for the Physics and Mathematics of the Universe, The University of Tokyo, Kashiwa, Japan 277-8583 (Kavli IPMU, WPI)}
\affiliation{Department of Astronomy, School of Science, The University of Tokyo, 7-3-1 Hongo, Bunkyo, Tokyo 113-0033, Japan}

\author{Xuheng Ding}
\affiliation{Kavli Institute for the Physics and Mathematics of the Universe, The University of Tokyo, Kashiwa, Japan 277-8583 (Kavli IPMU, WPI)}

\author{Michael A. Strauss}
\affiliation{Department of Astrophysical Sciences, Princeton University, 4 Ivy Lane, Princeton, NJ 08544, USA}

\author{Andy Goulding}
\affiliation{Department of Astrophysical Sciences, Princeton University, 4 Ivy Lane, Princeton, NJ 08544, USA}

\author{Malte Schramm}
\affiliation{Graduate school of Science and Engineering, Saitama Univ. 255 Shimo-Okubo, Sakura-ku, Saitama City, Saitama 338-8570, Japan}

\author{Hassen M. Yesuf}
\affiliation{Kavli Institute for the Physics and Mathematics of the Universe, The University of Tokyo, Kashiwa, Japan 277-8583 (Kavli IPMU, WPI)}
\affiliation{Kavli Institute for Astronomy and Astrophysics, Peking University, Beijing 100871, China}

\author{Mouyuan Sun}
\affiliation{Department of Astronomy, Xiamen University, Xiamen, Fujian 361005, China}

\author{Yongquan Xue}
\affiliation{CAS Key Laboratory for Research in Galaxies and Cosmology, Department of Astronomy, University of Science and Technology of China, Hefei 230026, China}
\affiliation{School of Astronomy and Space Science, University of Science and Technology of China, Hefei 230026, China}

\author{Simon Birrer}
\affiliation{Kavli Institute for Particle Astrophysics and Cosmology and Department of Physics, Stanford University, Stanford, CA 94305, USA}

\author{Jingjing Shi}
\affiliation{Kavli Institute for the Physics and Mathematics of the Universe, The University of Tokyo, Kashiwa, Japan 277-8583 (Kavli IPMU, WPI)}

\author{Yoshiki Toba}
\affiliation{Department of Astronomy, Kyoto University, Kitashirakawa-Oiwake-cho, Sakyo-ku, Kyoto 606-8502, Japan}
\affiliation{Academia Sinica Institute of Astronomy and Astrophysics, 11F of Astronomy-Mathematics Building, AS/NTU, No.1, Section 4, Roosevelt Road, Taipei 10617, Taiwan}
\affiliation{Research Center for Space and Cosmic Evolution, Ehime University, 2-5 Bunkyo-cho, Matsuyama, Ehime 790-8577, Japan}

\author{Tohru Nagao}
\affiliation{Research Center for Space and Cosmic Evolution, Ehime University, 2-5 Bunkyo-cho, Matsuyama, Ehime 790-8577, Japan}

\author{Masatoshi Imanishi}
\affiliation{National Astronomical Observatory of Japan, National Institutes of Natural Sciences (NINS), 2-21-1 Osawa, Mitaka,
Tokyo 181-8588, Japan}
\affiliation{Department of Astronomy, School of Science, The Graduate University for Advanced Studies, SOKENDAI, Mitaka,
Tokyo 181-8588}

\begin{abstract}
\noindent We measure the evolution of the \rel~relation using 584 uniformly-selected SDSS quasars at $0.2<z<0.8$. The black-hole masses (\mbh) are derived from the single-epoch virial mass estimator using the H$\beta$ emission line, and span the range $7.0<\logmbh/\msun<9.5$. The host-galaxy stellar masses (\m), which cover the interval $10.0<\logm/\msun<11.5$, are determined by performing two-dimensional quasar-host decomposition of the Hyper Suprime-Cam images and spectral energy distribution fitting. 
To quantify sample-selection biases and measurement uncertainties on the mass terms, a mock quasar sample is constructed to jointly constrain the redshift evolution of the \rel~relation and its intrinsic scatter (\sigmamu) through forward modeling.
We find that the level of evolution is degenerate with \sigmamu, such that both positive mild evolution (i.e, \mbh/\m~increases with redshift) with a small \sigmamu, and negative mild evolution with a larger \sigmamu~are consistent with our data.
The posterior distribution of \sigmamu~enables us to put a strong constraint on the intrinsic scatter of the \rel~relation, which has a best inference of $0.25_{-0.04}^{+0.03}$ dex, consistent with the local value. The redshift evolution of the \rel~relation relative to the local relation is constrained to be $(1+z)^{0.12_{-0.27}^{+0.28}}$, in agreement with no significant evolution since $z\sim0.8$. The tight and non-evolving \rel~relation is suggestive of a coupling through AGN feedback or/and a common gas supply at work, thus restricting the mass ratio of galaxies and their black holes to a limited range. Given the considerable stellar disk component, the \bulgerel~relation may evolve as previously seen at higher redshifts.\\
\end{abstract}

\section{Introduction}
\label{sec:intro}
It has been a puzzle for a couple of decades how the tight scaling relations, observed between the supermassive black hole (SMBH) mass and the properties of the bulge component of their host galaxies (e.g., bulge mass \mbulge, stellar velocity dispersion $\sigma_\star$), in the local universe are established \citep[e.g.,][]{Ferrarese2000, Gebhardt2000, Marconi2003, Haring2004, Kormendy2013}. The scenario in which the scaling relations result from a self-regulation procedure caused by AGN feedback has been widely proposed in theoretical models but still remains to be futher tested observationally \citep[e.g.,][]{Silk1998, King2003}. It is also debated whether SMBHs and galaxies coevolve in lockstep or one  predates the other. On the other hand, a non-causal origin through multiple merger-averaging has been proposed in which SMBHs and galaxies eventually converge to the local scaling relations via the central limit theorem \citep{Peng2007, Jahnke2011}.
The key clue to distinguishing these possibilities is likely imprinted in how the slope, normalization and intrinsic scatter of the scaling relations evolve with cosmic time.
Together, these quantities are governed by the growth history of galaxies such as through their merger history and the effect of AGN feedback. They can then be used as a diagnostic to constrain different galaxy evolution models \cite[e.g.,][]{Habouzit2020, Ding2020simu}. Unveiling the connections between SMBHs and galaxies thus requires accurate measurements of the scaling relations as a function of redshift and their dependencies on galaxy structural properties, in order to understand when and how these connections emerged. 

However, the scaling relations are still not well established beyond the local universe. Due to the difficulty of separating the bulge mass from the disk mass, many studies focus on the \rel~(i.e., the total stellar mass) relation instead.
The elevated $\mbh/M_*$ ratio at high redshifts ($z\sim0.5-2.5$) found by some studies \citep[e.g.,][]{Merloni2010, Decarli2010, Bennert2011b} suggests that BH growth predates that of galaxies. However, a deficient \citep{Ueda2018, Ishino2020} or  non-evolving \mbh/\m~ratio \citep[e.g.,][]{Shields2003, Jahnke2009, Cisternas2011, Schramm2013, Salviander2013, Sun2015, Matsuoka2015, Suh2020} has also been reported by many studies.  Thus, the issue has yet to be firmly settled. 

Many of the discrepancies can be explained by biases due to sample selection effects, which lead to an apparent evolution in the \rel~relation \citep{Treu2007, Lauer2007, Shen2010, Schulze2011, Schulze2014}. 
Specifically, considering that (1) the \rel~relation has an intrinsic scatter, (2) the stellar mass function (SMF) dramatically drops at the high-mass end and (3) luminous AGNs have on average more massive BHs, samples selected based on a
threshold in AGN luminosity will tend to select overmassive BHs with respect to stellar mass, and an AGN of a given \mbh~is more likely to be found in less massive galaxies which are more numerous \citep{Treu2007, Lauer2007}.
In addition, the measurements of BH mass beyond the local universe rely on broad-line AGN samples, whose BH masses are usually derived from the single-epoch virial method \citep[e.g.,][]{Kaspi2000, Vestergaard2002, Vestergaard2006}. \cite{Shen2010} pointed out that (see also \citealt{Shen2008, Shen2013}), as the virial \mbh~estimates depend on AGN luminosity (see Equation \ref{eq:mbh}), and there is non-negligible random scatter between instantaneous luminosity and emission-line width, the luminosity cut for a flux-limited AGN sample will cause the virial \mbh~to be statistically biased high (i.e., the luminosity-dependent bias in virial BH masses). Together, these effects lead to an apparent excess in $\mbh/\m$ for high-redshift luminous AGN samples. Moreover, \cite{Schulze2011} suggested that using AGN samples to trace the \rel~relation suffers a bias caused by a \mbh-dependent active fraction (or equivalently, the AGN duty cycle), in the sense that the observed \rel~relation for AGNs is biased towards low (high) BH masses if the active fraction decreases (increases) with increasing \mbh.

Recently, \cite{Ding2020a} studied a sample of 32 AGNs at $1.2<z<1.7$ with ACS + WFC3 imaging from the Hubble Space Telescope. After a detailed consideration of selection biases, they found that the evolution of the \rel~relation was perhaps mildly positive.
A surprising result from their study is that the intrinsic scatter appears to be similar to the local value. To confirm this with higher statistical significance,  larger AGN samples at high redshifts with reliable BH masses, stellar masses and galaxy structural properties are critically important. A careful treatment of sample selection biases is also needed.

In \cite{L21} (hereafter L21) we performed two-dimensional quasar-host galaxy decomposition for a sample of $\sim5000$ type~1 SDSS quasars \citep{Paris2018} at $0.2<z<1$, using deep  high-resolution images and accurate PSF models in the $grizy$ bands from the Subaru Strategic Program (SSP) survey \citep{Aihara2019} with Hyper Suprime-Cam \citep{Miyazaki2018}.
We detected the hosts in nearly all objects in the \hbox{\i-band} and measured properties of the host galaxies such as size, \sid, stellar mass and rest-frame colors. Our main finding was that quasar hosts are preferentially compact star-forming galaxies (SFGs). Thus we may be witnessing both active SMBH accretion and bulge growth triggered possibly by a compaction process \citep{Dekel2014, Zolotov2015, Silverman2019}. This large quasar sample is ideal for investigating not only the evolution of the normalization of the \rel~relation, but also its intrinsic scatter and its dependences on host-galaxy properties.

Therefore, this work makes use of the L21 sample to tackle this problem. This paper is organized as follows. In Section \ref{sec:sample} we describe our sample selection and methods to derive \m~and \mbh. 
The observed \rel~relation is presented and compared to that in the local universe in Section \ref{sec:observed}. We perform a detailed modeling of sample selection biases using mock AGN samples in Section \ref{sec:quantify_bias}, and present the intrinsic \rel~relation for our sample in Section \ref{sec:intrinsic}. The discussion and conclusions are presented in Sections \ref{sec:result} and \ref{sec:summary}, respectively. Throughout the paper we assume a flat $\Lambda$CDM cosmology with $\Omega_\Lambda$ = 0.7, $\Omega_{\rm m}$ = 0.3, and $H_0 = 70\rm\ km\ s^{-1}\ Mpc^{-1}$.  All magnitudes are given in the AB system. The masses and luminosities are given in units of \msun~and erg $\rm s^{-1}$, respectively.

\begin{figure*}
\includegraphics[width=0.49\linewidth]{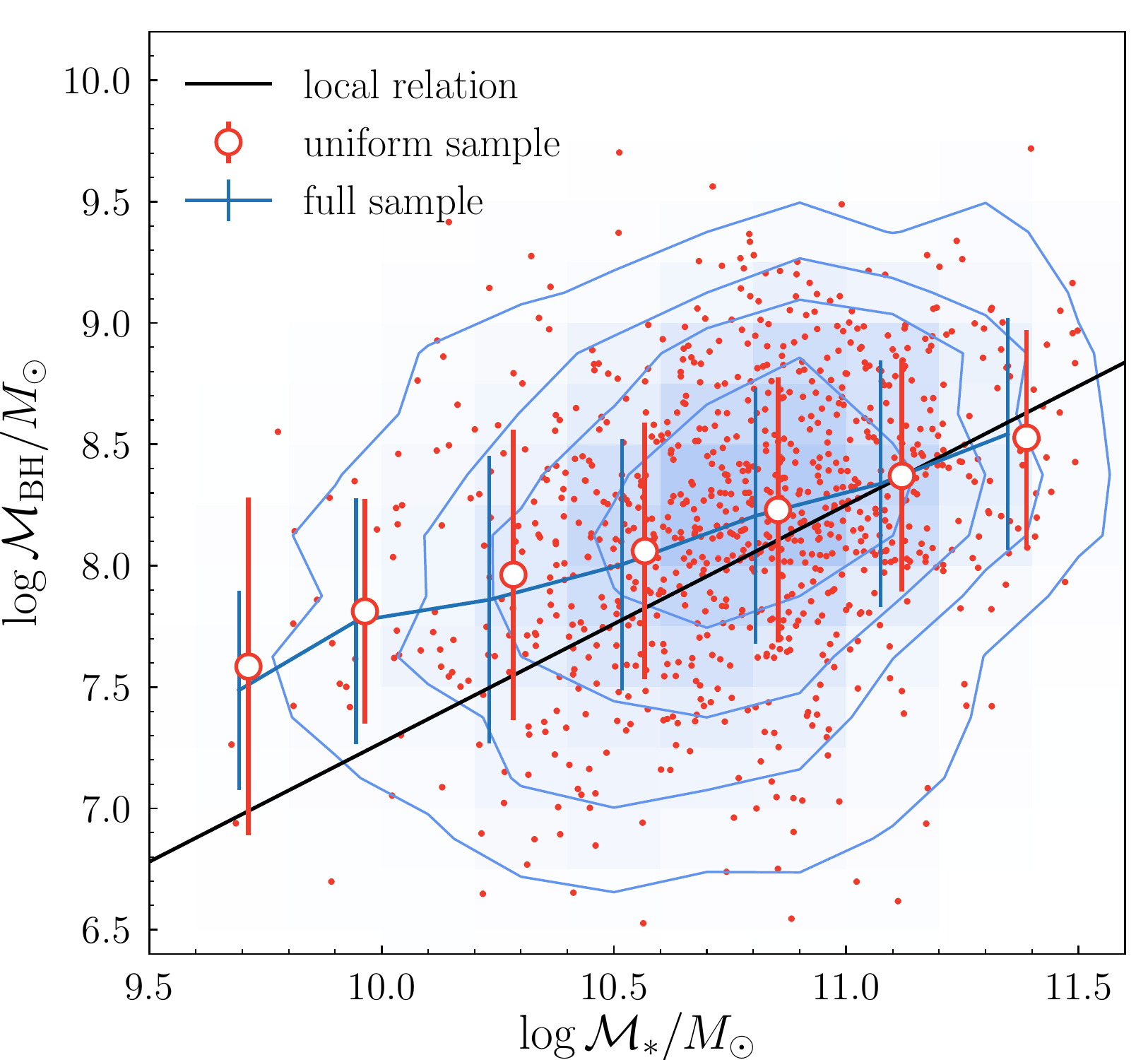}
\includegraphics[width=0.49\linewidth]{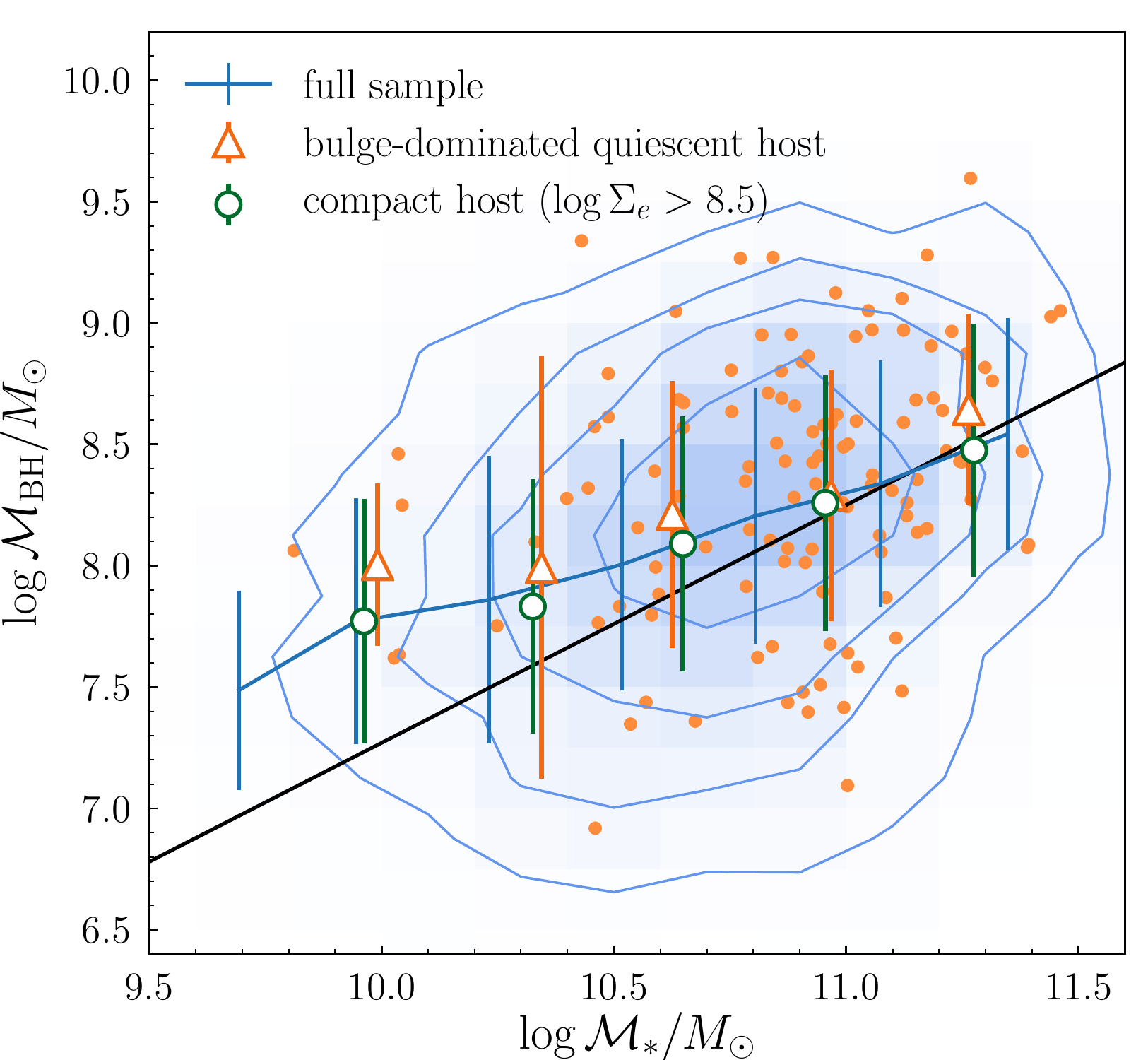}
\caption{Observed distribution and average \rel~relation for the SDSS-HSC quasar sample. The full sample is shown as the same blue contours and lines in both panels.  The left panel (Section \ref{sec:sample}) shows the uniform sample (red points and circles). The right panel (Section \ref{subsec:bulge}) shows quasars hosted by bulge-dominated quiescent galaxies (orange points and triangles) and compact galaxies (green circles), respectively.
Data points with error bars represent the mean and $1\sigma$ scatter in bins of \logm. The local relation compiled in \cite{Ding2020a} is shown as a black line. The observed \rel~distribution for SDSS quasars is affected by selection biases such that low-mass BHs are missing in our sample because of the luminosity limit.}
\label{fig:obs_mbh}
\end{figure*}

\section{Methods and Sample Selection}
\label{sec:sample}
In L21 we performed two-dimensional quasar-host image decomposition for a sample of $\sim5000$ type 1 quasars selected from the SDSS DR14 quasar catalog \citep{Paris2018} at $0.2<z<1$, using high-quality optical imaging data in $grizy$ bands from the HSC-wide survey. The HSC-wide survey has a $5\sigma$ point-source depth of $\sim$26 mag with a median seeing of 0.6$''$ in the \i-band.
The host galaxies and quasars are modelled as a two-dimensional \ss~profile and an unresolved point-source component characterized by the point spread function, respectively, using the state-of-the-art image modeling tool \lens~\citep{Birrer2015, Birrer2018}. The main outputs are the half-light radius, \sid~($n$) and decomposed galaxy and quasar fluxes. The superb spatial resolution of HSC images allowed us to accurately measure the host-galaxy fluxes in multiple bands and derive their stellar masses through spectral energy distribution fitting (SED; see Section 3.2 in L21) using CIGALE \citep{Boquien2019}. For the SED fits, we assume a \cite{Chabrier2003} initial mass function (IMF), a \cite{Bruzual2003} stellar population model, a delayed star-formation history and a \cite{Calzetti2000} extinction law. We decomposed simulated quasars and their host galaxies to verify the robustness of our stellar-mass measurements and correct for systematic effects. The typical error on \m~is estimated to be 0.2~dex based on the comparison with the stellar mass derived from the CANDELS dataset in the UV to mid-IR wavelengths \citep{Skelton2014}. We refer the readers to Section 4.2 in L21 for details.

L21 defined a final sample of 2424 quasars that passed further selection based on redshift limit ($z<0.8$), an assessment of the goodness-of-fit on image decomposition and SED fitting, and stellar mass limits ($\logm<11.5$ and $\logm > 9.3$, 9.8 and 10.3 at $z=0.3$, 0.5 and 0.7, respectively) thus having reliable stellar mass and structural measurements (see Section 5 in L21). Hereafter, we refer to these 2424 quasars as the full (i.e., parent) sample. 

The BH masses of these quasars are taken from \cite{Rakshit2020}, which are derived from the single-epoch virial estimator using the following expression:
\begin{equation}
\begin{split}
{\rm log} \Bigg(\frac{\mbh}{\msun} \Bigg) = A + B\, {\rm log}\,\Bigg(\frac{L_{\rm AGN}}{\,10^{44}\, \ergs} \Bigg)\\ +  2\,{\rm log}\,\Bigg(\frac{\nuw}{\rm km\ s^{-1}}\Bigg),
\label{eq:mbh}
\end{split}
\end{equation}
where $L_{\rm AGN}$ and \nuw~are the quasar continuum luminosity and the full width at half maximum of the broad emission line, respectively. We adopt the H$\beta$-based \mbh~as our fiducial value with $A=0.91$, $B=0.50$ and $L_{\rm AGN}=L_{5100}$ (\citealt{Vestergaard2006}; hereafter VP06). We note that using the {Mg~\scriptsize{II}}-based \mbh~at $z>0.5$ gives consistent results (the two BH masses have a mean offset of $0.01\pm0.37$~dex).
The typical uncertainty on virial \mbh~is $\sim0.4$~dex \citep[e.g.,][]{Vestergaard2006, Shen2013}.

This study requires an accurate characterization of the sample selection function (see Section \ref{subsec:bias}). However, SDSS quasars are selected using various criteria and are highly inhomogenous \citep{Paris2018}. To avoid any potential biases related to the selection of our quasars, we focus on 858 sources that are uniformly selected based on PSF-magnitude and optical color using single-epoch SDSS photometry. Specifically, we use the $ugri$ color-selected sample (225 sources) from SDSS I/II \citep{Richards2002}, and the CORE sample from SDSS BOSS (398 sources) and eBOSS (235 sources) surveys \citep{Ross2012, Myers2015} (hereafter the uniform sample). These samples are initially selected based on PSF-magnitude cuts of $15<i<19.1$ (for $ugri$) and $i>17.8$ \& $g$ or $r < 22.0$ (for CORE). Additional limits on SDSS colors (and WISE color only for the CORE sample) are used to identify quasar candidates. Figure \ref{fig:obs_mbh} plots the distribution of  quasars in the \rel~plane. The observed \rel~distribution for the uniform sample is similar to that of the full sample in terms of both the mean and the scatter, suggesting that the uniform sample is representative of the full sample. Even so, our main analyses presented below will be based on the uniform sample.

\section{The Observed \rel~relation}
\label{sec:observed}
{From the observed \rel~relation as shown in Figure \ref{fig:obs_mbh} (left panel), we confirm that type-1 quasars probe the massive end of the SMF and BH mass function (BHMF): the average \logm~and \logmbh~are 10.8 and 8.3, respectively, for the uniform sample. These quasars are thus likely the progenitors of the most massive BHs in the most massive, and predominantly bulge-dominated galaxies in the local universe, and are biased tracers of the underlying SMBH populations.

To investigate the evolution of the \rel~relation, we compare our observed result with the relation in the local universe in the form of $\logmbh = c_1\,\logm + c_2$ in Figure \ref{fig:obs_mbh}. 
We adopt the constants $c_1=0.97$ and $c_2=-2.44$ as given by \cite{Ding2020a}, which are derived by combining local quiescent galaxies from \cite{Haring2004} (HR04)\footnote{We adopt the HR04 relation since the VP06 virial estimator is calibrated based on that sample \citep{Tremaine2002}. To compare with the updated local relation given by \cite{Kormendy2013} (KH13), which has a +$\sim$0.3~dex offset relative to HR04, one needs to first recalibrate the virial BH mass based on their $\mbh-\sigma_\star$ relation, which is also offset by +$\sim$0.3 dex relative to the \cite{Tremaine2002} sample \citep{Onken2004}. That is, both the local \rel~relation and the BH mass for our quasars will increase by +$\sim$0.3 dex if we were to compare with KH13, leaving their relative offset largely unaffected.} and local AGN samples from \cite{Bennert2010, Bennert2011a}\footnote{We adopt the same VP06-based recipe to derive the H$\beta$-based BH mass as \cite{Ding2020a}. The BH mass for the local AGN samples in \cite{Bennert2010, Bennert2011a} have been recalibrated using the same recipe in \cite{Ding2020a} to derive the local baseline.}. Our quasars are typically located above the local relationship, especially for BHs in low-\m~galaxies, where the offset is as large as 0.5~dex. However, the SDSS survey is strongly biased to detect only the most luminous quasars with the highest \mbh~which drives the apparent evolution as demonstrated in Section \ref{sec:quantify_bias}. 

It is worth mentioning that our local baseline is actually the \bulgerel~relation (but $\mbulge \approx \m$ for the stellar mass range probed by our sample).
Specifically, the \cite{Haring2004} sample consists of bulge-dominated ($\mbulge \approx \m$) massive quiescent galaxies (mainly ellipticals and S0 galaxies), which were initially used to define the tight scaling relations between BHs and bulges in the local universe. Their bulge masses are derived from dynamical modeling, and range from $10.0 < \logmbulge < 11.5$. In addition, only the bulge masses for galaxies in the Bennert et al. AGN sample are used to construct the local baseline in \cite{Ding2020a}. The bulge masses in \cite{Bennert2010, Bennert2011a} are derived from AGN-bulge-disk decomposition using imaging data from the Hubble Space Telescope, which is similar to the method we adopted in L21 to derive stellar masses for the SDSS quasar sample. The same IMF and stellar population model are assumed in both works when converting the decomposed galaxy fluxes to bulge or stellar masses. Their bulge masses range from $9.0 < \logmbulge < 10.3$, and closely follow the \bulgerel~relation defined by  quiescent galaxies \citep{Bennert2011a, Bennert2021}. The inclusion of the local AGN sample thus helps to improve the constraint on the \bulgerel~relation at the low-mass end. In addition, the good consistency between the \bulgerel~relation for local AGNs and quiescent galaxies can be served as evidence that the bulge mass (and total stellar mass in our case) derived through image decomposition is comparable to those derived from dynamical modeling \citep{Padmanabhan2004}.
}

\section{Quantifying Sample-Selection Biases with Simulations}
\label{sec:quantify_bias}

\subsection{Method}
\label{subsec:bias}

\begin{figure*}
\includegraphics[width=\linewidth]{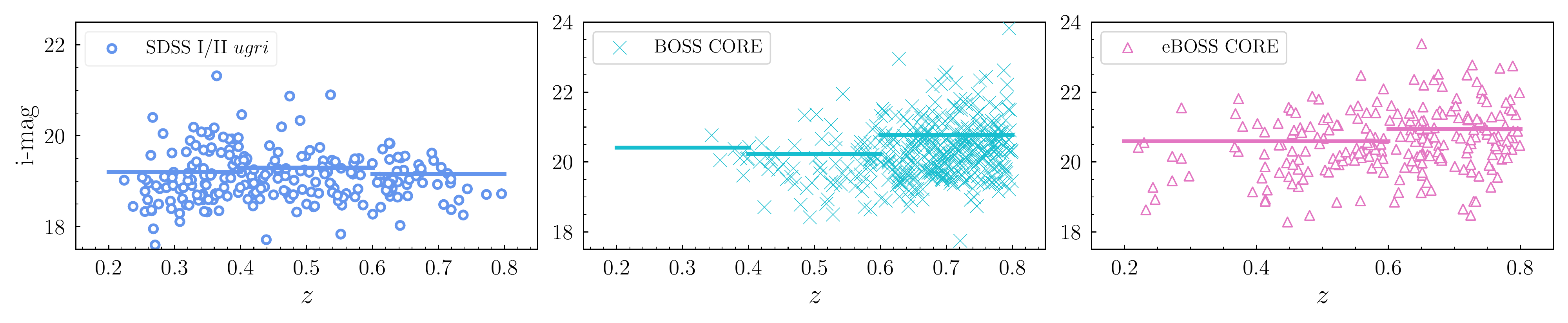}
\caption{The \i-band quasar-only magnitude as a function of redshift for the three observed quasar samples, shown separately. The magnitude cuts applied to both observed and simulated quasars are shown as solid horizontal lines. Objects lying below the segments are included in the subsequent analyses.}
\label{fig:mag_cut}
\end{figure*}

\begin{figure*}
\includegraphics[width=\linewidth]{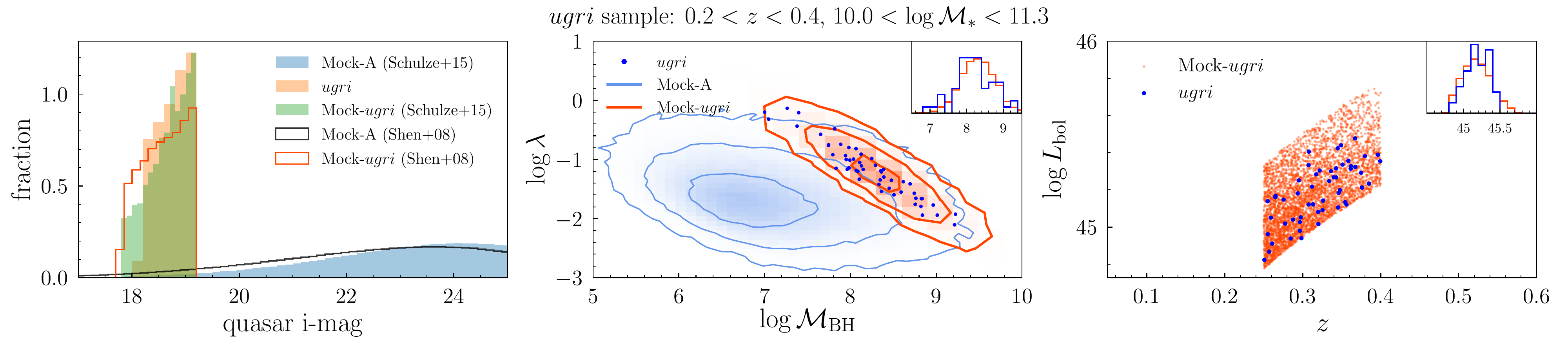}
\includegraphics[width=\linewidth]{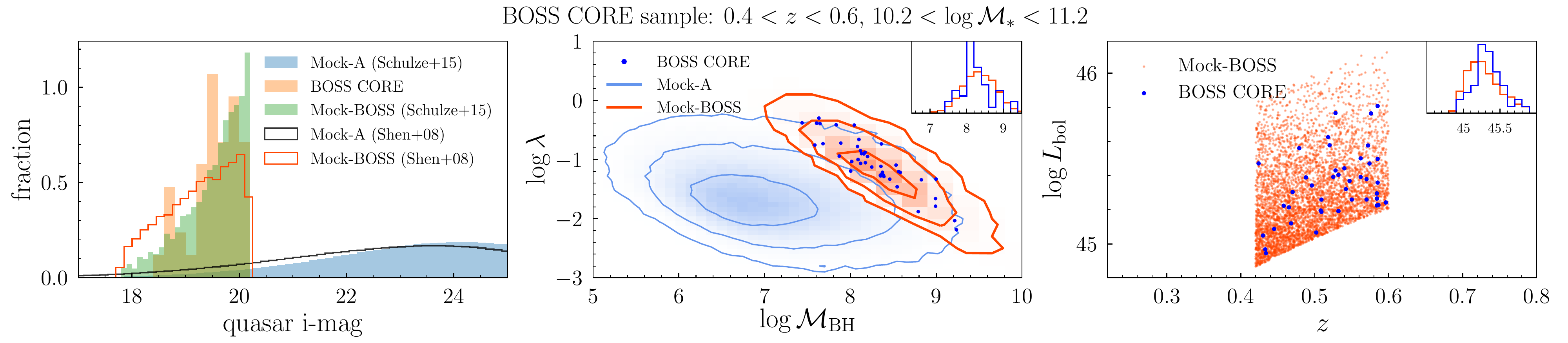}
\includegraphics[width=\linewidth]{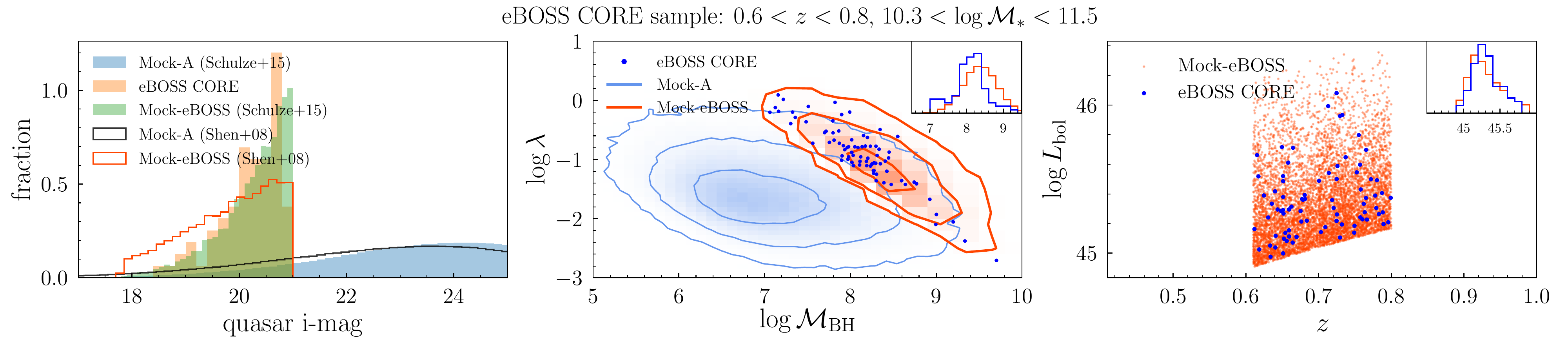}

\caption{Comparing simulated SMBH -- galaxy samples to each SDSS quasar sample. Mock-A represents the full simulated sample without selection effects being applied, while Mock-$ugri$, Mock-BOSS and Mock-eBOSS represent those with selection effects incorporated. The simulated and observed samples have matched redshift distributions. Left panel of the top row: distributions of the \i-band quasar-only magnitude at $0.2<z<0.4$ and $10.0 < \logm<11.3$ for the SDSS-$ugri$ sample and mock samples assuming different ERDs \citep{Shen2008, Schulze2015}. Middle and right panels of the top row: \edd~vs. \mbh~and \lbol~vs. $z$ in the same redshift and stellar-mass ranges, respectively. Eddington ratio for mock samples plotted here is derived from \lbol~and virial \mbh~instead of the initially sampled true \edd~in order to make a direct comparison with observations. The distributions of virial \mbh~(middle) and \lbol~(right) are plotted in the inset histograms. Middle row: similar to the top row but for the BOSS CORE sample at $0.4<z<0.6$ and $10.2<\logm<11.2$. Bottom row: similar to the top row but for the eBOSS CORE sample at $0.6<z<0.8$ and $10.3<\logm<11.5$.}
\label{fig:mock}
\end{figure*}

As introduced in Section \ref{sec:intro}, the observed \rel~relation for high-redshift AGN samples is affected by various selection biases, which can lead to an apparent offset in \mbh/\m~as shown in Figure \ref{fig:obs_mbh}. We use Monte Carlo simulations to build a mock AGN sample which incorporates, as closely as possible, the selection of SDSS quasars in order to account for such effects on the observed \rel~relation and to derive its intrinsic evolution.

In practice, our simulation starts with the SMF, the local mass-scaling relation between BHs and their host galaxies and an evolutionary model to generate mock galaxies and SMBHs over a range of redshifts. We relate the simulated {\it true} BH mass to observables such as {\it virial} BH mass, emission line width and AGN luminosity through Equation \ref{eq:mbh} and an intrinsic Eddington ratio distribution (ERD). We then apply selection criteria on AGN luminosity and emission line width to account for selection biases, and then compare the resulting \rel~relation with observations to generate the likelihood for different model parameters (i.e., evolution, dispersion).

Specifically, we randomly sample the galaxy stellar mass function (SMF) at $z=0.15-0.85$ (assuming a flat redshift distribution) provided by \cite{Muzzin2013} to build a sample of simulated galaxies. The SMF is formulated as:
\begin{equation}
\Phi(s) = {\rm ln10}\, \Phi_* 10^{(s-s_*)/(1+\alpha)} \times  {\rm exp}(-10^{s-s_*}),
\label{eq:mass_function}
\end{equation}
where $s \equiv \logm$ and the values of $\Phi_*$, $s_*$ and $\alpha$ are 12.16, 11.22, $-$1.29 at $z<0.5$ and 16.25, 11.00 and $-$1.17 at $z>0.5$, respectively. 

We then assign a BH mass for each simulated galaxy from the \rel~relation. To account for possible evolution, we assume an evolutionary model for the mean \rel~relation formulated as  
\begin{equation}
\deltalogmbh = \gamma\, {\rm log}\,(1+z),
\label{eq:evolution}
\end{equation}
where $\deltalogmbh \equiv \logmbh - (c_1\,\logm - c_2)$ is the deviation of the observed BH mass from that predicted by the local relation. This model is anchored to the HR04 relation at $z=0$ at the massive end. Thus in this work, we are investigating how our quasars with massive BHs and  host galaxies may evolve to bulge-dominated, massive quiescent galaxies such as those found in the HR04 sample, settling on the tight local scaling relations.

Putting these model components together, the distribution of $m \equiv \logmbh$ at a given $s$ is parameterized as:
\begin{equation}
\Phi(m|s) = \frac{1}{\sqrt{2\pi}\sigmamu}\, {\rm exp}\Bigg(-\frac{(m - (c_1 s + c_2 + \gm\,{\rm log}\,(1+z)))^2}{2\sigmamu^2}\Bigg),
\label{eq:mbh_m_prob}
\end{equation}
assuming that the intrinsic scatter \sigmamu~follows a normal distribution. Thus for a given distribution of $s$ as sampled from Equation \ref{eq:mass_function} and a grid of parameters (\gm, \sigmamu), we randomly select {\it true} BH masses ($m_{\rm true}$) from Equation \ref{eq:mbh_m_prob} to create a mock distribution of SMBHs at each redshift. 

The selection biases on \mbh~are mainly related to our observational cuts on AGN luminosity  \citep{Lauer2007, Shen2010, Schulze2014}. 
To assign bolometric luminosities to our mock SMBHs, we use the intrinsic ERD of broad-line AGNs -- which has the form of a Schechter function \citep{Schechter1976} -- as given in \cite{Schulze2015}\footnote{The \mgii-based BH masses in \cite{Schulze2015} are calibrated using the formula from \cite{Shen2011} which matches the VP06 H$\beta$ masses on average.}:
\begin{equation}
    \phi(\lambda) = \frac{\phi^*}{\lambda_*} \Big(\frac{\lambda}{\lambda_*}\Big)^{\alpha_{\lambda}}\,{\rm exp}\Big(-\frac{\lambda}{\lambda_*}\Big),
\end{equation}
where the redshift-dependent values of $\phi^*, {\rm log}\,\lambda_*, \alpha_\lambda$ are given in Table 1 of \cite{Schulze2015} derived from a uniformly selected SDSS DR7 quasar sample \citep{Schneider2010}. We randomly sample the \edd~function as given above and combine it with the sampled $m_{\rm true}$ distribution to derive the corresponding \lbol. We then convert this into the $i$-band magnitude ($m_i$) assuming a bolometric correction to the rest-frame \i-band luminosity of 12 \citep{Richards2006}; the $k$-correction to the observed frame is made by assuming that the UV-to-optical quasar SED follows a power-law with $\alpha_\nu = -0.44$ \citep{VandenBerk2001}. 

The parameters generated from the above steps are considered as the {\it{true}} properties for the mock sample. We then add uncertainties to those parameters to generate observables that are comparable to real observations. We first add scatter to each sampled $s$ by randomly generating a stellar mass from a Gaussian with a mean of $s$ and a standard deviation of $0.2$ dex (see Section 4.2 in L21).
For each simulated SMBH, we also derive a virial \mbh~to include measurement uncertainties on the true \mbh. 
We assume that \nuw~follows a log-normal distribution with the mean value calculated by inserting the sampled $m_{\rm true}$ and \lbol~in Equation \ref{eq:mbh} with a scatter of $\nuw=0.17$~dex \citep{Shen2008}. 
The sampled distributions of \nuw~and \lbol~are then reinserted to Equation \ref{eq:mbh} to derive the distribution of virial BH mass \mvir. By doing so, the resulting scatter of \mvir~to $m_{\rm true}$ is $\sim 0.34$~dex, and the variation of luminosity at a fixed true \mbh~is compensated by line width. As a result, our virial \mbh~is an unbiased estimator of the true \mbh~irrespective of AGN luminosity. Therefore, we are ignoring the possible luminosity-dependent virial \mbh~biases in \cite{Shen2010}, which is a suitable assumption for our relatively low-$z$ and low-luminosity ($\overline{{\rm log}\,L_{5100}}=44.4$) sample based on \hb~\citep{Shen2012, Shen2013, Wang2020}. The impact of the possible existence of this bias will be discussed in Section \ref{subsec:error}. 

Note that the previous steps created a sample of SMBHs that are all actively accreting as observed BLAGNs.  This contains an unstated assumption that the \rel~relation traced by AGNs (without the luminosity limit) is the same as that of the underlying galaxy population. However, the two relations could be intrinsically different if the AGN duty cycle depends on \mbh~\citep{Schulze2011, Schulze2015}, meaning that SMBHs with different \mbh~have different probabilities to be observed as AGNs, thus biasing the observed \rel~relation. We will further discuss this issue in Section \ref{subsec:error}.

\begin{figure*}
\includegraphics[width=\linewidth]{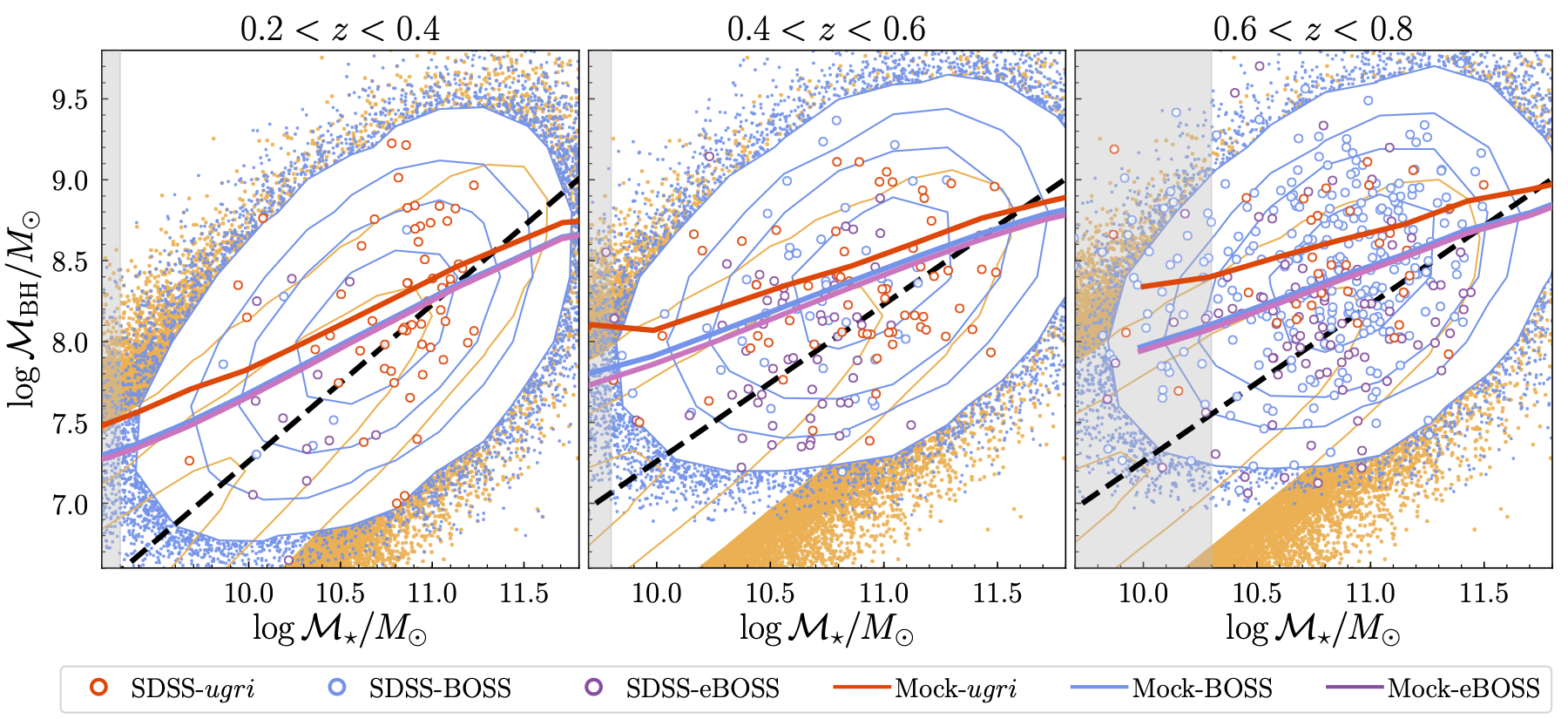}
\caption{Distributions and average \rel~relations in three redshift intervals. The observed SDSS $ugri$, BOSS and eBOSS samples are plotted as red, blue and purple empty circles, respectively.
The orange contours ($1\sigma-3\sigma$ levels) represent the Mock-A sample (no selection effects incorporated) using true \mbh~and \m~values, while the small orange points represent the \mbh~and \m~both with noise added.
The solid curves represent the average \rel~relation for the Mock-B sample with various selections applied as described in the text (Mock-$ugri$: red; Mock-BOSS: blue; Mock-eBOSS: magenta).
The blue contours and small blue points show the distribution of the Mock-BOSS sample as an example to illustrate the impact of selection effects.
The black dashed line represents the local relation compiled in \cite{Ding2020a}. The gray shaded region marks our stellar-mass cut at the median redshift of each redshift interval. 
}
\label{fig:bias}
\end{figure*}

\subsection{Mock Catalogs}
\label{subsec:simu_result}

Combining each sampling step with different input parameters for evolution and intrinsic scatter (\gm, \sigmamu), we  generate a mock AGN catalog with known true $\mbh$, virial $\mbh$, $\m$, $\edd$, $\lbol$, $m_i$ and redshift.
We create $5\times10^7$ mock AGNs for each set of assumed values of \gm~and $\sigmamu$, and apply the selection criteria used for our uniform SDSS quasar (i.e., observed) sample. Specifically, we apply a cut at $\nuw > 1000\ \kms$ to include SMBHs identified as a BLAGN. For the magnitude limit, we do not use the initial SDSS PSF-magnitude limits (i.e., $15<i<19.1$ or $i>17.8\ \&\ g\,{or}\,r<22$). Rather, we derive an effective magnitude limit $m_i$ for each SDSS quasar using \lbol~(assuming the same $k$-correction as simulated objects), and set a redshift-dependent limit at 90\% of the peak of the $m_i$ distribution.  

The magnitude limits are set in this way for two reasons. First, unlike the simulated quasars, the SDSS PSF-magnitude still includes host-galaxy emission. Second, SDSS samples are incomplete at faint magnitudes with a significant drop in number density (Figure 6 in \citealt{Paris2018}). Therefore, the simulated sample will contain many more faint quasars (thus low-mass BHs) than observed if we set the cut at e.g., $i\lesssim22$, for the CORE sample, leading to an underestimation of the selection bias. We therefore apply a strict pure-quasar magnitude cut to both observations and simulations to artificially create a magnitude-matched sample.
In principle, the $k$-correction used to derive $m_i$ for SDSS quasars could vary from source to source \citep{Richards2001}. However, our aim here is to derive an average magnitude limit. Thus, we adopt the same correction values for both simulated and observed samples. The final magnitude cuts for each subsample as a function of redshift are shown in Figure~\ref{fig:mag_cut}. We also set a bright cut at $m_i>17.8$ to mimic the quasar selection for the BOSS and eBOSS surveys. This leaves a final sample of 584 sources that are used in the following analyses. 

We define the full mock sample (without applying any selection cuts) as Mock-A, and the mock sample which shares the same sample-selection criteria as our SDSS quasars as Mock-B  (consisting of Mock-$ugri$, Mock-BOSS, and Mock-eBOSS samples). 
In the first row of Figure \ref{fig:mock}, we compare the distributions of quasar magnitude, the \edd~vs. \mbh~plane, and the \lbol~vs. $z$ plane for an example Mock-$ugri$ sample ($\gm=0.0$, $\sigmamu=0.3$) along with the observed SDSS $ugri$ sample at $0.2<z<0.4$ and $10.0<\logm<11.3$ to illustrate how well our simulated samples reproduce the observed distributions. 
Similar comparisons for the BOSS CORE sample with $0.4<z<0.6$ and $10.2<\logm<11.2$, and for the eBOSS CORE sample with $0.6<z<0.8$ and $10.3<\logm<11.5$ are plotted in the second and third rows in Figure \ref{fig:mock}, respectively. After applying the selection cuts, the distributions of Mock-B samples are in good agreement with observations, thus validating our fiducial model assumptions (see Section \ref{subsec:error} for the discussion on other model choices).

In Figure~\ref{fig:bias} we show the \rel~relation for one of  our simulations having $\gamma=0.0$ and $\sigmamu=0.3$, corresponding to a non-evolving \rel~relation with the intrinsic scatter equal to the local value \citep{Haring2004, Kormendy2013}. SDSS quasars are overlaid as empty circles.  It can be seen that the observed scatter for simulated SMBHs is much larger than the assumed intrinsic value (comparing the blue and orange contours): the maximum perpendicular spread can be as high as $\sim1.5$~dex due to large uncertainties associated with virial BH mass and stellar mass estimates.

With our cuts on quasar magnitudes and emission-line widths, our Mock-B samples are strongly affected by selection biases, which is most significant for less massive galaxies \citep{Shen2015}. This can be seen by comparing the orange points and blue contours in Figure~\ref{fig:bias} which indicate the simulated sample before and after the selection effects are incorporated. As a result, only relatively massive BHs in massive galaxies are retained in the sample, as indicated by the hard lower limit on \mbh, and the large offsets between the solid curves (with selection effects) and the black dashed line (no selection effects). 

Considering the stellar mass range probed by our quasar sample, the average offsets as we approach our lower stellar mass limit can be as large as 1.0 dex for Mock-$ugri$ (which has a higher luminosity threshold), and 0.5 dex for Mock-BOSS and Mock-eBOSS. 
At the massive end, the average \rel~relation for Mock-B samples lies below the local relation, which can be tentatively seen for the observed SDSS quasars in Figure~\ref{fig:obs_mbh}. Due to the rapid drop-off of the SMF and a non-negligible error in the stellar mass estimates, there are more low-\m~galaxies hosting low-mass BHs that scatter into the high-\m~bins than the opposite, leading to a decrement in the average \mbh~at the high-\m~end. The bright magnitude limit ($m_i > 17.8$) also slightly contributes to the negative \deltalogmbh~as the most luminous (massive) BHs are excluded from Mock-B samples.

Overall, the distribution of SDSS quasars on the $\m-\mbh$~plane is in good agreement with the Mock-B sample. The biased \rel~relation for the mock samples generally follows the observed data points up to $z\sim0.8$, suggesting that selection biases can largely account for the observed offsets between SDSS quasars and the local relation, and that measurement uncertainties contribute significantly to the observed scatter for high-redshift quasar samples.

\section{The Intrinsic \rel~Relation} 
\label{sec:intrinsic}

\begin{figure*}
\includegraphics[width=\linewidth]{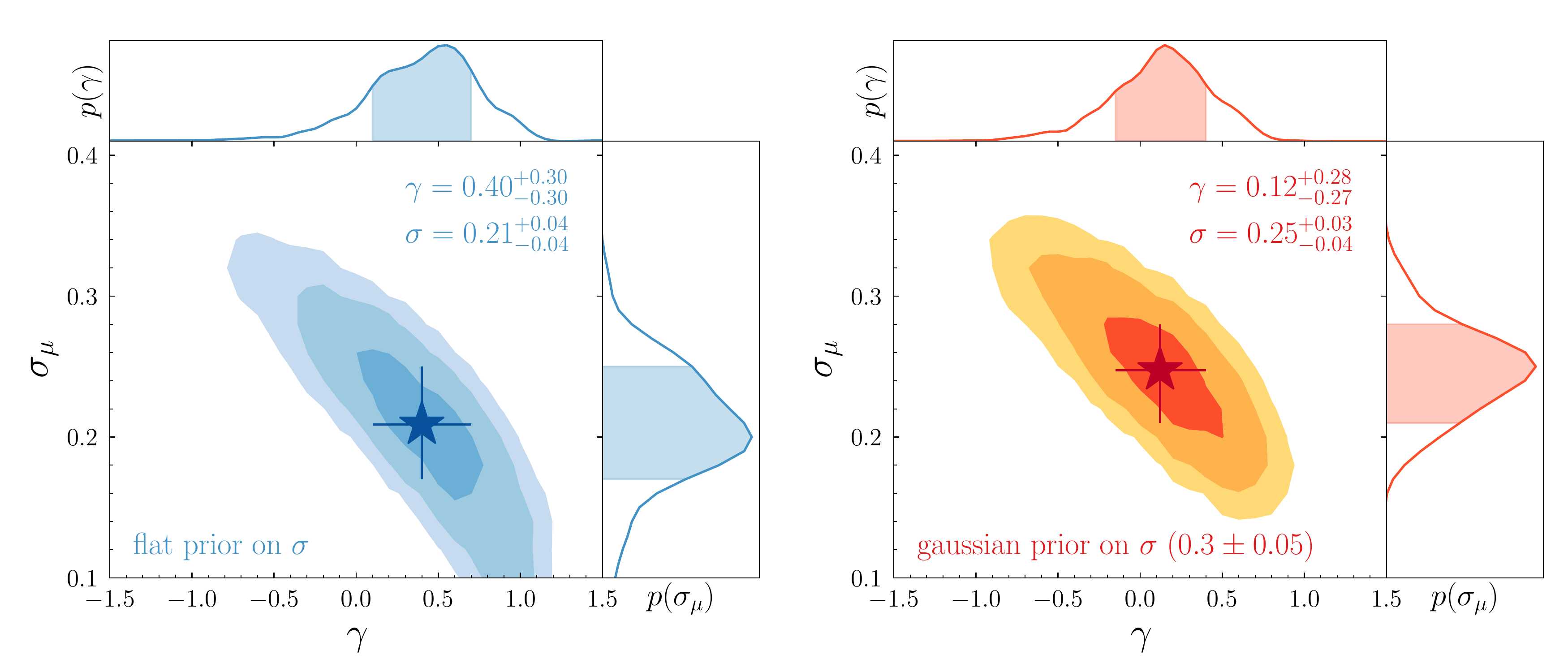}

\caption{Constraining the evolution factor \gm~and intrinsic scatter \sigmamu~of the \rel~relation (Equation \ref{eq:evolution}), assuming a flat prior (left panel) or a gaussian prior (right panel) on \sigmamu, based on a Monte Carlo simulation. The colored areas indicate the 68\%, 95\%, and 99\%~confidence regions of the posterior probability. The histograms are the integrated one-dimensional posterior distributions of \gm~and \sigmamu. The distributions have been slightly smoothed for visual purposes. The stars indicate our best inferences with $1\sigma$ error bars, which suggest a non-evolving \rel~relation out to $z\sim0.8$, particularly in the case with a gaussian prior.}
\label{fig:post}
\end{figure*}

We constrain the evolutionary parameters by maximizing the total probability of observing each \mbh~for SDSS quasars at a given \m~using samples from Mock-B with different combinations of \gm~and \sigmamu~\citep[e.g.,][]{Bennert2010, Park2015}.
This is done by selecting simulated objects having similar stellar mass ($\Delta \logm<0.2$) and redshift ($\Delta z<0.05$) to each observed quasar, generating a distribution in simulated virial \mbh~for the selected objects, and then computing the likelihood. The latter is derived by convolving the aforementioned mock \mbh~distribution with the probability density distribution of each observed \mbh, which is assumed to be a Gaussian with its standard deviation equal to the $1\sigma$ measurement error\footnote{This error only reflects the spectral fitting uncertainty on AGN luminosity and line width. The median error for our sample is 0.1 dex.} given by \cite{Rakshit2020}.

The posterior distributions of \gm~and \sigmamu~assuming an uninformative flat prior on the intrinsic scatter are shown in Figure \ref{fig:post} (left panel). There is a degeneracy between \gm~and \sigmamu: both  a  positive \gm~(i.e, \mbh/\m~increases with redshift) with a small \sigmamu, or a negative \gm~with a larger \sigmamu~are consistent with our data. Our best inferences are $\gm=0.40_{-0.30}^{+0.30}$ and $\sigmamu=0.21_{-0.04}^{+0.04}$. Assuming that the intrinsic scatter is not significantly smaller than the local value, we adopt a gaussian prior on \sigmamu~with a mean of 0.3 (i.e., the intrinsic scatter for the local sample) and a standard deviation of 0.05. This gives  $\gm=0.12_{-0.27}^{+0.28}$ and $\sigmamu=0.25_{-0.04}^{+0.03}$ (see the right panel in Figure \ref{fig:post}).  In both cases, our results suggest that the \rel~relation does not evolve significantly (no more than $\sim0.2$ dex considering the $1\sigma$ uncertainty range) since $z\sim0.8$. 

Interestingly, we find that the intrinsic scatter is not larger than the local value assuming either a flat or a gaussian prior \citep[e.g.,][]{Ding2020a}.
This statement is also true when separating our sample into higher and lower mass galaxies based on the average stellar mass ($\overline{\logm}=10.8$). The best-fit values are  $\gm=0.13_{-0.28}^{+0.32}$, $\sigmamu=0.21_{-0.04}^{+0.04}$ for massive galaxies, and $\gm=0.04_{-0.54}^{+0.56}$, $\sigmamu=0.29_{-0.05}^{+0.04}$ for less massive galaxies (assuming gaussian priors)\footnote{We adopt the values assuming gaussian priors as our fiducial result for both the full sample and each subsamples, since it is unlikely that the intrinsic scatter for massive galaxies can be as small as $\sigmamu \sim 0.13$ with $\gm \sim 0.54$, as is the case when assuming a flat prior. For less massive galaxies, assuming a flat prior gives consistent results ($\gm\sim0.18$, $\sigmamu \sim 0.27$) as adopting a gaussian prior ($\gm\sim0.04$, $\sigmamu \sim 0.29$).}. Note that a different $\gm$ for each \m~bin would indicate that the intrinsic offsets in \deltalogmbh~for massive and less massive galaxies are different, i.e., the slope of the \rel~relation evolves, and galaxies having different stellar masses may evolve through different pathways. However, we are not able to robustly distinguish \gm~for low- and high-mass galaxies because of significant error on \gm. Whether the slope of the \rel~relation evolves is yet to be further explored.

\section{Discussion}
\label{sec:result}

\subsection{\rel~Relation and its Evolution}
\label{sec:discuss_intrinsic}

\begin{figure*}
\includegraphics[width=\linewidth]{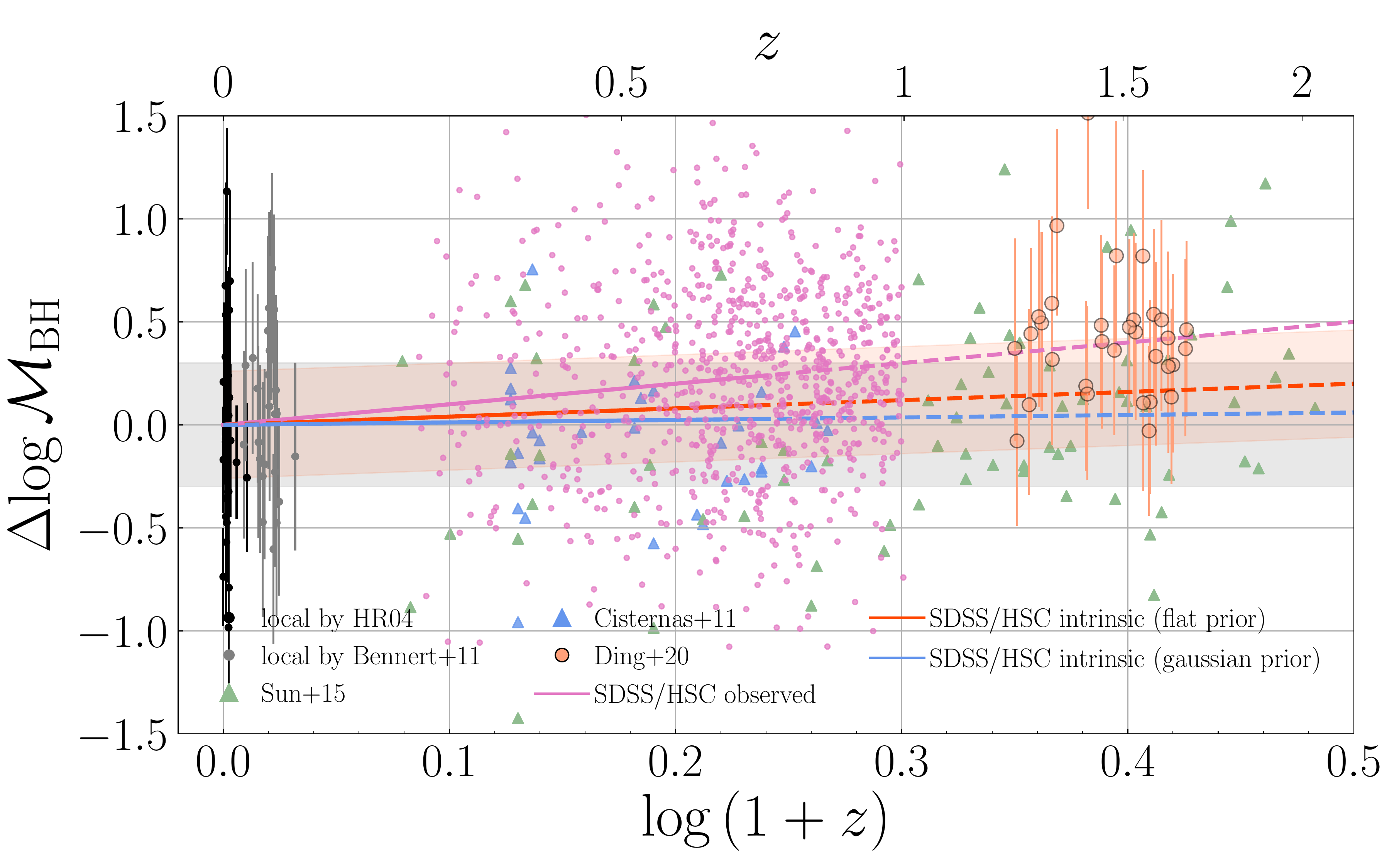}
\caption{Evolution of \deltalogmbh~as a function of redshift. Individual SDSS quasars as well as their observed mean offsets relative to the local relation are shown as pink points and the pink solid ($z<0.8$)--dashed ($z>0.8$) line, respectively. The red (blue) solid-dashed line represents the intrinsic evolution of SDSS quasars after accounting for selection biases assuming a flat (gaussian) prior on \sigmamu, with its best-fit value shown by the red shaded region.
Local objects from \cite{Haring2004} and \cite{Bennert2010, Bennert2011a}, as well as  high-redshift studies which reported a constant \citep{Cisternas2011, Sun2015} or a slightly-evolving \citep{Ding2020a} \mbh/\m~are plotted for comparison. The gray shaded region marks the scatter for the local sample. The typical uncertainty for high-redshift \mbh~estimates is $0.4$~dex, which is only shown for the \cite{Ding2020a} sample for simplicity.}
\label{fig:bias_evo}
\end{figure*}

\begin{table}
\renewcommand{\arraystretch}{1.2}
\caption{Model fitting results in the form of Equation~\ref{eq:evolution} to the observed offset (\deltalogmbh) as a function of redshift.}
\centering
\begin{tabular}{ccc}
\hline
\hline
 & \gm & \sigmamu \\
\hline
observed & $1.00\pm0.07$ & 0.48\\
intrinsic (flat prior) & $0.40_{-0.30}^{+0.30}$ & $0.21_{-0.04}^{+0.04}$\\
intrinsic (gaussian prior) & $0.12_{-0.27}^{+0.28}$ & $0.25_{-0.04}^{+0.03}$\\
\hline
\end{tabular}
\label{table:params}
\end{table}

In Figure \ref{fig:bias_evo} we plot the observed values of \deltalogmbh~(Equation \ref{eq:evolution}) for SDSS quasars and the model fits including the intrinsic relation (i.e., after accounting for selection biases) as a function of redshift. The fitting parameters are summarized in \hbox{Table \ref{table:params}}.
SDSS quasars at $0.8<z<1.0$ in our parent sample are also plotted only to illustrate the large observed offset at higher redshifts. AGN samples from previous studies at $0.1<z<2$ that reported a constant \citep{Cisternas2011, Sun2015} or a slightly-evolving \citep{Ding2020a} \mbh/\m~are also shown. {The IMF and virial factors assumed to derive \m~and \mbh~in these works are the same as ours.} The main result here is that our bias-corrected \rel~relation for SDSS quasars does not evolve significantly since $z\sim0.8$, and that the intrinsic scatter is similar to the local value. 

The similar scatter of the \rel~relation for SDSS quasars to that of the local relation suggests that merger averaging is not the dominant mechanism for establishing the tight scaling relations. A similar result has been reported by \cite{Ding2020a} at $z\sim1.5$ but with less statistical significance.
Under the merger scenarios, the scatter is expected to be redshift-dependent \citep{Peng2007, Jahnke2009, Hirschmann2010}. 
Based on a pure-merging model in a large scale cosmological simulation, \cite{Hirschmann2010} found that the intrinsic scatter decreases by $\sim0.05$~dex from $z\sim1.5$ to $z\sim0.8$, and further decreases by $\sim0.05$~dex and converges to the local value at $z\sim0$ as a result of successive minor/major mergers.
This is not supported by our observations since the measured \sigmamu~($0.25_{-0.04}^{+0.03}$), particularly for quasars in less massive ($\logm<10.8$) galaxies ($\sigmamu=0.29_{-0.05}^{+0.04}$), is not larger than the local value. The disk-like morphology for our quasar hosts (see Section 6.3 in L21) and their generally low asymmetry values (Yesuf et al. in prep) also suggest that the majority of them have not experienced recent violent major mergers. Therefore, our result rules out that the tight scaling relation is purely stochastic in nature driven by multiple random mergers, and suggests that a physical coupling between SMBHs and their hosts is likely at work to make galaxies settle to a restricted range of \mbh/\m. This could be due to AGN feedback which self-regulates BH accretion and star formation \citep[e.g.,][]{DiMatteo2005}, or it may be driven by a common cold gas reservoir which provides fuel for both SMBH and galaxy growth \citep[e.g.,][]{Shangguan2020}.

With that being said, the fact that \sigmamu~is lower ($0.21^{+0.04}_{-0.04}$; see Section \ref{subsec:error}) in massive galaxies than in less massive galaxies ($\sigmamu=0.29_{-0.05}^{+0.04}$) suggests that, hierachial assembly of \mbh~and \m~through mergers may indeed help to reduce the scatter of the scaling relation \citep{Peng2007, Hirschmann2010, Jahnke2011}, as massive galaxies may have experienced more mergers.  

We do recognize that our sample is limited in redshift range. It does not probe the peak epoch of merging activity which is likely at $z>1.5$ \citep[e.g.,][]{Husko2021}. Therefore, the relative importance of merger averaging, gas fueling and AGN feedback in establishing the scaling relations is not conclusive. Their contributions could also vary with cosmic time. Moreover, the era of peak BH accretion and star formation has passed, especially for massive systems. Thus, the physical processes that built the tight \rel~relation observed at present are mostly close to completion. This imposes requirements on similar analyses for large AGN samples at higher redshifts in order to constrain the \rel~relation at early cosmic times.

It will also be enlightening to compare our observations with cosmological simulations by applying the same selection cuts to simulated SMBHs (as was done in \citealt{Ding2020simu}). \cite{Habouzit2020} have analyzed the evolution of the \rel~relation in the cosmological simulations Illustris, TNG 100, TNG 300, Horizon-AGN, EAGLE and SIMBA,  which adopt different initial conditions and physical recipes such as AGN feedback mode and efficiency in galaxy evolution models. The median of the \rel~relations for massive galaxies in all simulations appears to evolve little since $z\sim1$. However, the intrinsic scatter behaves differently in different simulations. For example, the scatter in the Horizon-AGN simulation remains small ($\sim0.2$~dex) and does not change with redshift and stellar mass. However, the scatter in the EAGLE simulation increases significantly ($>0.5$~dex) toward lower stellar masses and higher redshifts. The variation of the scatter in different simulations is mainly set by the strength of supernovae and AGN feedback, as well as the relevant role of gas accretion and mergers in driving SMBH and galaxy growth. Additional effort is thus required to refine the sub-grid physics in simulations and reproduce scaling relations that better match with our observations.

\subsection{Bulge-dominated Quasar Host Galaxies}
\label{subsec:bulge}
As mentioned in Section \ref{sec:observed}, we are  comparing the \mbh/\m~ratio to the \mbh/\mbulge~ratio in the local universe. As a result, our quasars follow the local scaling relation within the $1\sigma$ uncertainty, but with their total stellar mass rather than the bulge mass \citep{Jahnke2011, Cisternas2011, Sun2015}. Combining the results presented in this work and L21, we now discuss how our quasars may evolve and settle on the tight \bulgerel~relationship in the local universe.

In Figure \ref{fig:obs_mbh} (right panel) we show the \rel~relation for quasars with bulge-dominated quiescent host galaxies. These galaxies are those with \ss~indices $>3$ (see Section 6.3 in L21) and red $U-V$ color (see Section 6.4 in L21).
It appears that they follow the same \rel~relation as the whole sample. This is also true when considering the most compact objects in our sample (right panel in Figure \ref{fig:obs_mbh}) as selected by their sizes or surface mass densities (e.g., ${\rm log}\,\Sigma_e\, (M_\odot/{\rm kpc^2}) > 8.5$; see Section 6.6 in L21), which we can measure at  higher precisions than the \sid.
Since these centrally-concentrated galaxies suffer the same sample-selection biases, their intrinsic \mbh/\m~distributions should be the same as the entire  quasar sample, thus they should also follow the local relation\footnote{Their best-fit evolutionary parameters are consistent with that of the entire sample within $1\sigma$ uncertainties, albeit with much larger errors.}. 

These quasars are likely the progenitors of massive ellipticals in the local universe, with their current $\mbh/\m\, (\approx\mbulge)$ value being already consistent with the local $\mbh/\mbulge$ ratio \citep{Schramm2013, LJ2021}. Their subsequent assembly is likely driven by dry minor mergers \citep{Naab2009, vandokkum2010}.
Since the \mbh/\m~ratio of galaxies with $\logm\approx10$ appears similar to more massive galaxies at these redshifts, and only a small amount of mass is added to both quantities when merging with even smaller galaxies, a similar mass ratio can be maintained.

Considering bulge-dominated hosts yet to be fully quenched, their mass ratio will evolve as
\begin{equation*}
\frac{\mbh(t_0)}{\mbulge(t_0)} = \frac{\mbh(t_1)+{\int_{t_1}^{t_0} {\rm BHAR}(t)dt}}{\mbulge(t_1)+{\int_{t_1}^{t_0} {\rm SFR}(t)dt}},
\end{equation*}
where BHAR and SFR are the black-hole accretion rate and star-formation rate, respectively.
To maintain a constant \mbh/\mbulge~ratio from $z=0.8$ to $z=0$, a constant $\rm BHAR/SFR \approx \mbh(t_0)/\mbulge(t_0)$ is required, which is seen in observations \citep{Mullaney2012, Yang2019}. This result suggests that for bulge-dominated massive galaxies, the \bulgerel~relation is already established at $z\sim0.8$.

\subsection{Disk-dominated Quasar Host Galaxies}
\label{subsec:disk}

On the other hand, the majority of our quasars are found to be hosted by compact star-forming galaxies with substantial disks (see Section 6 in L21). The peak ($n\sim1$) and median \ss~indices~($n\sim2$) for our sample suggest that most quasars have a bulge-to-total (B/T) ratio smaller than $\sim50\%$ (see Figure 11 in \citealt{Ding2020a}). They may be building up their bulges through a compaction process \citep{Dekel2014, Zolotov2015, Lapiner2021} as they are on average more compact than \m- and $z$-matched inactive star-forming galaxies \citep{Silverman2019, Molina2021}. Since it is the total stellar mass of these quasars that follows the local \bulgerel~relation, their current bulge masses are thus likely to be deficient by a factor of $\gtrsim2$ at a given \mbh~relative to the local relation \citep{Peng2006, Treu2007, Jahnke2009, Bennert2011b, Cisternas2011, Schramm2013, Ding2020a}.

In light of a simple argument that local SMBHs with $\logmbh>8.0$ are almost all hosted by ellipticals and bulge-dominated ($\overline{\rm B/T}\sim80\%$) S0 galaxies \citep{Kormendy2013}, we know that, independent of the exact evolutionary pathways, our massive quasars ($\overline{\logmbh}=8.3$) which are still actively accumulating their BH masses, will eventually end up as early type galaxies at $z\sim0$. 
For these quasars to settle on the local \bulgerel~relation, their hosts must experience a structural change to evolve from being disk-dominated to bulge-dominated, while the mechanism(s) which leads to such morphological transformation must be more efficient in building up the bulge while limiting the fueling of the SMBH. This could be achieved through a redistribution of disk stars to the bulge \citep{Jahnke2009, Cisternas2011, Sun2015, Ding2020a}, which can increase the bulge mass while maintaining a relatively constant \mbh/\m~ratio. It is not clear which mechanism (or mechanisms) causes the disk-to-bulge transition, with likely candidates
including minor mergers, major mergers, or a dynamical compaction process caused by the accretion of counter-rotating streams or/and internal disk instabilities \citep{Hopkins2010, Dekel2014, Zolotov2015}. At the end phase of such a transformation, the bulge mass catches up to the BH mass, after which they coevolve with each other as we discussed previously for bulge-dominated galaxies. 

Overall, our results show that \mbh~is better correlated with the total galaxy mass in the redshift and mass ranges probed by our sample (\citealt{Jahnke2009, Schramm2013, Sun2015};  see \citealt{Yang2019} for an alternative conclusion). The \bulgerel~relation seen in the local universe may thus be a relic of a tight \rel~relation resulting from correlated SMBH and galaxy growth at higher redshifts, as \m~will eventually transfer from disks to bulges as discussed before. Our quasars may happen to be in such a structural transition phase as detailed in L21. We emphasize that our result is only valid for massive systems. Low-mass BHs (e.g., $\logmbh\sim7$ at $z=0$) may evolve in a different way (due to e.g., insufficient AGN feedback power) and follow a different \rel~relation in the local universe \citep[e.g.,][]{Reines2015, Davis2018}.

\subsection{Dependence of the \rel~Relation on Galaxy Structural Properties}
As discussed in Section \ref{subsec:bulge}, SDSS quasars with bulge-dominated or compact host galaxies appear to follow the same \rel~relation as the rest of the sample. This is equivalent to the finding in L21 (see their Section 6.5) that the galaxy size--stellar mass relations for quasars hosts with different \mbh~are indistinguishable, at least for the \mbh~range probed by our sample. 
As detailed in L21, these quasars may be building up their central mass concentration through nuclear star formation or a redistribution of disk stars to the bulge. Since the latter process does not affect BH mass, the indistinguishable \rel~relation for compact and extended hosts would suggest that, for quasars which are rapidly building their bulges, the same mechanism which triggers gas inflow and active nuclear star formation tends to fuel SMBHs at similarly higher rates \citep{Trump2013} than those in extended galaxies.

\subsection{Impact of Model Assumptions and the Limitations of the Current Sample}
\label{subsec:error}
The aforementioned results inevitably suffer from various uncertainties, which depend on the accuracy of our measured \mbh~and \m, as well as model assumptions. Here we briefly discuss how different model choices may affect our results.

Adopting a different SMF, such as that given by \cite{Davidzon2017} or that of SFGs only in \cite{Muzzin2013} has little impact on our result. In addition, we are ignoring the biases caused by detecting BLAGNs which requires a high quasar-to-host contrast, and reliably measuring stellar mass which requires a high host-to-quasar contrast. However, such biases are expected to be small \citep{Sun2015} and cancel each other to some degree as they act in the opposite directions. 

Moreover, when generating mock AGN samples (as done in Section \ref{subsec:bias}), we assumed that the \rel~relation for AGNs (without the luminosity limit) and the underlying galaxy populations are the same. However, \cite{Schulze2015} suggested that the active SMBH fraction ($f_{\rm active}$), derived by comparing the active BHMF traced by type-1 quasars to the total BHMF scaled from the galaxy SMF (as an approximation of duty cycle of type 1 AGNs), strongly decreases with increasing \mbh~at $z<1$.
As a result, massive BHs have little probability to be observed as AGNs, which could bias the observed scaling relation in an opposite way relative to the \cite{Lauer2007} bias. On the other hand, the duty cycle of AGNs (refer to the whole AGN population rather than type 1s) can also be derived by modeling the continuity equation. Using this technique, \cite{Marconi2004} found that the duty cycle at $z<3$ is nearly constant at $\logmbh>7.5$, thus having negligible impact on our result. However, the duty cycle presented in \cite{Aversa2015} increases with \mbh~over $z=0-6$, thus further biasing the observed \rel~relation towards massive BHs. With the uncertainties in mind, we estimate the impact of a \mbh-dependent duty cycle to our result by assigning each mock source a weight in the sampling procedure. The weight is determined by the duty cycle for each source given their true \mbh, representing the possibility of each simulated SMBHs being observed as active SMBHs. An \edd~value drawn from the \cite{Schulze2015} ERD is then assigned to the selected active BHs. The subsequent sampling steps are the same as described in Section \ref{sec:quantify_bias}. As a result, when adopting the \cite{Schulze2015} duty cycle, the average \rel~relation (solid curves in Figure \ref{fig:bias}) for mock AGN samples (with selection effects) decreases by $\sim0.1$~dex. This causes the best-fit \gm~to increase by $\sim0.2$ compared to our fiducial model. However, if we were to adopt the \cite{Aversa2015} duty cycle instead, the best-fit \gm~decreases by $\sim0.2$. The best-fit \sigmamu~remains the same when adopting different models of AGN duty cycle ($|\Delta \sigmamu|<0.02$).

In addition, we ignore the possible luminosity-dependent bias in virial \mbh. This assumption likely has little impact at $z<0.5$ for \hb-based \mbh~estimates. However, a possibly weak but non-zero bias is inferred by \cite{Shen2012} at $z>0.5$; although, it is currently unclear how strong it is. If non-negligible, the average virial \mbh~for our quasars in the highest redshift bin might be slightly overestimated ($\sim0.1$~dex; \citealt{Shen2010}). If taking this into account, the remaining mild positive offset seen in Figure \ref{fig:bias_evo} could be completely erased.

Another influential assumption in our result is the  intrinsic ERD for broad-line AGNs \citep{Shen2008, Kelly2013, Schulze2015}. 
Since the detection limit of \mbh~decreases with increasing \edd~at a fixed luminosity threshold, low-mass BHs can thus survive our luminosity cut if the ERD is higher in the mean, leading to a stronger contribution from intrinsic positive evolution to the observed \deltalogmbh.
As a reference, adopting the log-normal ERD presented in \cite{Shen2008} which is higher by $\sim$0.2--0.6 dex (depends on \mbh) in the mean than that of \cite{Schulze2015} results in $\gm \sim 1.0$ and $\sigmamu \sim 0.26$. However, the best-fit model using the \cite{Shen2008} ERD produces more bright AGNs (see Figure \ref{fig:mock}) than are observed, especially at $\logm>10.8$ where the \i-band magnitude distribution is nearly flat. 
On the other hand, adopting the \cite{Kelly2013} ERD for which the number density of ${\rm log}\,\edd < -1.0$ sources can be 1--2 orders of magnitude higher than that of \cite{Schulze2015} results in too few bright sources. Therefore, we adopt the \cite{Schulze2015} ERD as our fiducial model.
It is worth mentioning that studies based on X-ray selected AGN samples have found that the ERD (traced by $L_{\rm X}/\m$ assuming a constant \mbh/\m~ratio) depends on galaxy properties, such as stellar mass and galaxy color \citep[e.g.,][]{Aird2018}. Therefore, assuming a universal ERD over the entire \m~range in this work is likely to be an oversimplification. This, together with the effect of a possible \mbh-dependent duty cycle (which is not considered in Figure \ref{fig:mock}), could be responsible for the aforementioned discrepancies in the quasar magnitude distributions using different ERDs. However, there could be substantial differences in sample properties (e.g., obscuration) between optically and X-ray selected AGNs. It is thus beyond the scope of this work to fully explore the possible functional forms of the ERD in the literature. 

The best-fit \sigmamu~is mainly determined by the uncertainties being added to the mass terms for mock AGNs (0.2 dex for \m~and 0.34 dex for \mbh). Adopting larger uncertainties will further increase the spread of simulated SMBHs on the \rel~plane, thus the resulting \sigmamu~will be even smaller, leaving our main conclusions unaffected. On the other hand, it is unlikely that our \m~and \mbh~can be constrained to be much more precise than the uncertainties we assumed. Therefore, \sigmamu~should not be significantly underestimated.

Besides the limitations on model assumptions, our SDSS-HSC quasar sample is by no means complete. The most luminous (e.g., $\loglbol>46$) SDSS quasars which are likely to have the highest \mbh~are missed in the HSC survey (see Figure 1 in L21). This could explain the lack of sources having $\logmbh>9.2$ at the massive end in Figure \ref{fig:bias}, and could be partly responsible for the small scatter of our high-\m~subsample (Section \ref{sec:intrinsic}). However, the missing objects are not expected to have a significant impact on our result as they are intrinsically rare, and the result for the low-\m~subsample is less affected by this issue.
In addition, each quasar selection algorithm imposes color cuts, and is designed to target different quasar populations (e.g., the BOSS survey is designed to mainly select quasars at $z>2.2$). A detailed modeling of how the color cuts may bias the observed \mbh~distribution is beyond the scope of this paper. Therefore, we assumed that the combination of the $ugri$, BOSS and eBOSS samples is able to fill in the observed \mbh-\m~plane above certain luminosity limits, and only focused on the biases induced by the magnitude limits.

It is now clear that the observed \rel~relation depends on a complex interplay between the intrinsic evolution, the underlying population properties (e.g., SMF, ERD, AGN duty cycle) and the sample-selection function.
To reduce the model-dependency would require not only a large but also a well-defined distant quasar sample. Namely, the sample needs to have a uniform and well-controlled selection function, and accurate measurements of both SMBH (e.g., \mbh, ERD) and galaxy (e.g., \m~and SMF) properties that are consistently derived as local galaxies. These requirements are only partly achieved by this study. Our work thus does not focus on deriving an accurate evolutionary parameter, but mainly aims to show that under reasonable assumptions, the significant apparent excess in \mbh/\m~for high-redshift quasars can be mostly explained by selection effects, and we see no statistically significant evidence for evolution in the \rel~relation since $z\sim0.8$.   Fortunately, the aforementioned issues have little impact on the derived intrinsic scatter. Thus our result of a similar scatter to the local value should be robust.

Due to the error of \gm~and the existence of systematic uncertainties as mentioned above, we cannot rule out a mild positive/negative evolution with the current data set. A uniformly-selected quasar sample which is $\sim4-5$ times larger in size would reduce the statistical error on \gm~to $<0.1$. This can be achieved when the entire HSC survey is completed. Although great effort is required to deal with systematic uncertainties which will become significant once the statistical error is reduced. 

It is also critically important to push the current studies to the peak epoch of AGN activity, star-formation and merging activity, as discussed in Section \ref{sec:discuss_intrinsic}. Furthermore, with the Atacama Large Millimeter/submillimeter Array, it is now possible to probe the $\mbh-\mathcal{M}_{\rm dyn}$ relation (using dynamical mass) of quasars even out to $z\sim7$ \citep{Onoue2019, Izumi2019, Izumi2021}, and we may soon be able to directly map the starlight from these quasars with the James Webb Space Telescope \citep{Marshall2021}.

\section{Concluding Remarks}
\label{sec:summary}
We study the evolution of the \rel~relation for 584 uniformly-selected SDSS quasars at $0.2<z<0.8$. Our sample spans the range $10.0<\logm<11.5$ and $7.0<\logmbh<9.5$, and are biased towards the massive end of the SMF ($\overline{\logm}=10.8$) and BHMF ($\overline{\logmbh}=8.3$). Assuming an evolutionary model (Equation \ref{eq:evolution}) with different intrinsic scatter (Equation \ref{eq:mbh_m_prob}), we generate mock AGN samples which mimic the measurement uncertainties and sample-selection function of observations in order to make a direct comparison with the observed \rel~distribution of SDSS quasars (Figure \ref{fig:bias}). As a result, we recover the intrinsic \rel~relation free of observational biases through forward modeling (Figures \ref{fig:post} and \ref{fig:bias_evo}).

We find that the bias-corrected \rel~relation traced by SDSS quasars does not evolve significantly since $z\sim0.8$ (\deltalogmbh~no more than $\sim\pm0.2$~dex considering the $1\sigma$ uncertainty). In addition, the intrinsic scatter ($\sigma_\mu=0.25_{-0.04}^{+0.03}$ dex) is consistent with the local value. Combined with the fact that the majority of quasar hosts have disk-like morphology, our results suggest that a physical connection between SMBH and galaxy growth possibly through AGN feedback and/or a common gas supply is likely at work to maintain a tight and non-evolving mass scaling relation. 
Mergers may further help tighten the scaling relation, since \sigmamu~for massive galaxies ($\logm>10.8$) is smaller than that of less massive galaxies ($\sigmamu = 0.21^{+0.04}_{-0.04}$~dex vs. $0.29^{+0.04}_{-0.05}$~dex).

The \mbh/\m~ratio for our quasars is consistent with the \mbh/\mbulge~ratio of the local sample (consisting of massive, bulge-dominated quiescent galaxies), which is true for both bulge-dominated ($\mbulge\approx\m$) and disk-dominated quasars. 
As a result, the \bulgerel~relation for bulge-dominated quasars is already established at $z\sim0.8$ and is maintained to $z\sim0$.
The disk-dominated quasars also follow the local \bulgerel~relation, but with their total stellar mass rather than the bulge mass. Their bulges need to increase in mass by a factor of $\gtrsim2$ in the subsequent $\sim 3-7$~Gyrs for them to align onto the local \bulgerel~relation. This could be achieved through a morphological transition event (due to e.g., mergers, internal disk instabilities) which redistributes the disk stars to the bulge while maintaining a relatively constant \mbh/\m~ratio.

These results prefer a scenario where \mbh~is better correlated with \m~than \mbulge~within the redshift and mass ranges examined. The \bulgerel~relation discovered in the local universe may be a relic of a tight \rel~relation established at earlier cosmic times, as \m~will eventually transfer to \mbulge~for massive galaxies hosting massive BHs. Our quasars may be experiencing such a structural transformation phase as described in a companion paper \citep{L21}.

\acknowledgments 
We thank the referee for valuable suggestions that helped to improve the manuscript.
We thank Tommaso Treu, Knud Jahnke and Guang Yang for helpful discussions and suggestions.
J.Y.L. acknowledges support from the China Scholarship Council. J.D.S. is supported by the JSPS KAKENHI Grant Number JP18H01251, and the World Premier International Research Center Initiative (WPI Initiative), MEXT, Japan. J.Y.L. and Y.Q.X. acknowledge support from the National Natural Science Foundation of China (NSFC-12025303, 11890693, 11421303), the CAS Frontier Science Key Research Program (QYZDJ-SSW-SLH006), the K.C. Wong Education Foundation and the science research grants from the China Manned Space Project with NO. CMS-CSST-2021-A06. M.Y.S. acknowledges support from the National Natural Science Foundation of China (NSFC-11973002).

\bibliographystyle{aasjournal}

\end{document}